\documentclass[aps,twocolumn]{revtex4}
\usepackage{color}
\usepackage{epsfig}
\begin{document}

\title{Local anisotropy and giant enhancement of local electromagnetic
  fields in fractal aggregates of metal nanoparticles}

\author{Sergei V. Karpov}
\affiliation{L.V. Kirensky Institute of Physics, Russian Academy of
  Sciences, Siberian Branch, Krasnoyarsk 660036, Russia}

\author{Valeriy S. Gerasimov and Ivan L. Isaev}
\affiliation{Department of Physics and Engineering, Krasnoyarsk State
  Technical University, Krasnoyarsk 660028, Russia}

\author{Vadim A. Markel}
\affiliation{Departments of Radiology and Bioengineering,
  University of Pennsylvania, Philadelphia, PA 19104}

\begin{abstract}
  We have shown within the quasistatic approximation that the giant
  fluctuations of local electromagnetic field in random fractal
  aggregates of silver nanospheres are strongly correlated with a
  local anisotropy factor $S$ which is defined in this paper. The
  latter is a purely geometrical parameter which characterizes the
  deviation of local environment of a given nanosphere in an aggregate
  from spherical symmetry. Therefore, it is possible to predict the
  sites with anomalously large local fields in an aggregate without
  explicitly solving the electromagnetic problem.  We have also
  demonstrated that the average (over nanospheres) value of $S$ does
  not depend noticeably on the fractal dimension $D$, except when $D$
  approaches the trivial limit $D=3$.  In this case, as one can
  expect, the average local environment becomes spherically
  symmetrical and $S$ approaches zero.  This corresponds to the
  well-known fact that in trivial aggregates fluctuations of local
  electromagnetic fields are much weaker than in fractal aggregates.
  Thus, we find that, within the quasistatics, the large-scale
  geometry does not have a significant impact on local electromagnetic
  responses in nanoaggregates in a wide range of fractal dimensions.
  However, this prediction is expected to be not correct in aggregates
  which are sufficiently large for the intermediate- and
  radiation-zone interaction of individual nanospheres to become
  important.
\end{abstract}

\date{\today}
\maketitle

\section{Introduction}
\label{sec:intro}

Electromagnetic properties of fractal nanostructures have continuously
attracted attention since the late 1980-ies due to their rather
unusual physical properties and the possibility of numerous
applications, as described in several reviews of the
subject~\cite{kreibig_book_95,shalaev_96_2,stockman_98_1,shalaev_book_00,karpov_book_03,roldugin}.
Of special interest are aggregates of metal nanoparticles in hydrosols
and percolation clusters (metal-dielectric composites) which have, in
particular, exhibited the effects of giant enhancement of
nonlinear-optical
responses~\cite{butenko_90_1,stockman_92_1,zhuravlev_92_1,shalaev_96_1},
inhomogeneous localization of electromagnetic
eigenmodes~\cite{stockman_96_1,stockman_97_1}, and optical
memory~\cite{karpov_88_1,safonov_98_1,markel_99_3,kim_99_1,bragg_01_1}.

Theoretical description of the electromagnetic responses of disordered
fractal aggregates has been closely interconnected with numerical
simulations. This is due to the fact that a fully analytic solution
to the problem of interaction of an electromagnetic field with a large
random fractal aggregate has not been devised. Some approximate
theoretical approaches were based on the first Born~\cite{martin_87_1}
and mean-field~\cite{berry_86_1} approximations, approximations based
on few-body interaction (binary~\cite{markel_90_1,markel_91_1} or
binary-ternary~\cite{stockman_97_1} approximations), and various
phenomenological scaling
laws~\cite{stockman_92_1,markel_90_1,markel_91_1,stockman_95_1,stockman_91_1,shalaev_92_1}.
The first Born and the mean-field approximations are not, generally,
useful in the spectral regions where excitation is resonant. While
off-resonant electromagnetic properties of fractal aggregates are of
interest in atmospheric
physics~\cite{mountain_88_1,andreev_95_1,mikhailov_96_1}, the research
in electromagnetics of metal fractal aggregates is primarily focused
on resonant interactions. The few-body approximations and the scaling
laws proved to be very useful for qualitative theoretical description
at the early stages of research.  However, increasingly more realistic
simulations revealed that these approaches do not provide quantitative
results. Currently, they are effectively obsolete. A brief overview of
the progression of numerical models used to simulate electromagnetic
responses of fractal aggregates is given in the next paragraph.

The theoretical and computational description has been primarily based
on a model of an aggregate of $N$ touching identical spherical
nanoparticles. Each nanoparticle, and sometimes the aggregate as a
whole, are assumed to be much smaller in size than the external
wavelength. (Polydisperse aggregates built from spheres of different
size have also been recently
addressed~\cite{karpov_00_1,perminov_04_1,rautian_04_1}.) In order for
an aggregate to be considered fractal, the number of primary spheres,
must be large, typically, $\gtrsim 10^3$. Initially, simulations were
based on the dipole approximation.  In this approximation, each sphere
is assigned a dipole moment located at its center. The spheres then
interact with each other and the external field via dipole radiation
fields as described by $3N$ coupled-dipole
equations~\cite{markel_91_1}. In the late 1980-ies and early 1990-ies,
numerical solution of dense linear systems of $\gtrsim 10^3$ equations
was a difficult computational task.  Therefore a model of {\it
  diluted} aggregates was adopted and used, for example, in
Refs.~\cite{stockman_92_1,markel_91_1,stockman_95_1,shalaev_91_1,shalaev_92_2,shalaev_93_1,shalaev_94_2}.
According to this model, an aggregate of $N$ touching spheres (where
$N$ can be very large) is {\it diluted}, i.e., spheres were randomly
removed from the aggregate with the probability $1-p$, where $p\ll 1$.
Then the coordinates of the remaining spheres are rescaled according
to ${\bf r} \rightarrow p^{1/D}{\bf r}$, where $D$ is the fractal
dimension. This procedure does not change the density-density
correlation function of the aggregate in some intermediate region.
However, it does change the local structure of the aggregate
substantially. The few-body approximations and scaling laws were
largely validated with the model of diluted aggregates.  However, when
computations with non-diluted clusters became feasible, it was found
that both the few-body approximations and the scaling laws are
inaccurate~\cite{markel_96_1}. The deviation from the scaling laws has
been explained by the phenomenon of inhomogeneous
localization~\cite{stockman_98_1}; however, the theoretical relation
of this phenomena to the aggregate geometry has not been clarified.
Additionally, it has been well known that account of excitation of
higher multipole modes is important for touching nanoparticles, even
when the size of each nanoparticle is much smaller than the external
wavelength~\cite{sansonetti_80_1,gerardy_80_1,claro_82_1,mackowski_95_1}.
In particular, the dipole approximation failed to properly describe
experimentally observed red shifts in extinction spectra of colloid
aggregates~\cite{markel_96_1,danilova_93_1}. To remediate this
problem, a phenomenological model of geometrical renormalization have
been introduced~\cite{markel_96_1,markel_01_1} and, recently,
computations beyond the dipole approximation have been
performed~\cite{markel_04_3}.

The combination of findings contained in the above-cited references
strongly suggest that the local structure of aggregates is of primary
importance.  However, the local structure of random fractal
nanoaggregates has not been so far the focus of research. In this
paper we, for the first time, point to a strong correlation between
the anisotropy of local environment and enhancement of local field in
fractal aggregates within the {\it quasistatic approximation}. In
particular, we find that the correlation coefficient of the local
anisotropy factor $S$ (introduced below) and the value of a local
squared dipole moment $\vert {\bf d} \vert^2$ can be as high as $0.75$
and tends to grow with the wavelength. We have found that the average
local anisotropy factor is almost independent of fractal dimension
in the range $1.7<D<2.8$. Note that this result is expected to change
in large aggregates where intermediate- and far-zone interaction is
important.

The paper is organized as follows. In Section~\ref{sec:lap_def} the
local anisotropy factor is introduced. The dependence of the local
anisotropy factor on the fractal dimension of aggregates and other
parameters for computer-simulated fractals is discussed in
Section~\ref{sec:sim}. Section~\ref{sec:opt} contains results
concerning the correlation of local electromagnetic fields and the
local anisotropy factor.  The electromagnetic calculations in this
section were performed with the method of coupled
multipoles~\cite{mackowski_95_1,markel_04_3}, e.g., without the dipole
approximation. Finally, Section~\ref{sec:sum} contains a summary of
obtained results.

\section{Definition of the local anisotropy factor}
\label{sec:lap_def}

The definition of local anisotropy factor introduced in this paper is
based on an analogy with ellipsoids. An ellipsoid is a geometrical
object that can exhibit either perfect spherical symmetry, or strong
anisotropy, depending on its eccentricity.

Consider a general ellipsoid excited by a linearly polarized
monochromatic external wave of amplitude ${\bf E}_0$. In the
quasistatic limit, the polarization ${\bf P}$ inside the ellipsoid is
independent of position and can be found from

\begin{equation}
\label{Eq_1} 4\pi \left[ {1 \over {\epsilon - 1}} + \left( {1
\over 3} - \hat{Q}
  \right) \right]  {\bf P} =  {\bf E}_0 \ ,
\end{equation}

\noindent
where the tensor $\hat{Q}$ is given by

\begin{equation}
\hat{Q} = \int_V \hat{G}_0(0, {\bf r}^{\prime}) d^3 r^{\prime}
\ .
\end{equation}

\noindent
Here $\hat{G}_0({\bf r}, {\bf r}^{\prime})$ is the regular part of the
quasistatic free-space dyadic Green's function for the electric field.
The integral is taken over the volume of the ellipsoid, $V$, and is
independent of position. Therefore, it is evaluated at the center of
ellipsoid, ${\bf r}=0$. A unique property of ellipsoids is that
$\hat{Q}$ is diagonal in the reference frame whose axes are collinear
to the main axes of the ellipsoid. Correspondingly, if $E_{0,\alpha}$
are the Cartesian components of the external electric field in the
same reference frame, the solution to (\ref{Eq_1}) is

\begin{equation}
\label{Eq_2} P_\alpha = {{E_{0,\alpha}} \over
{4\pi\left[1/(\epsilon-1) +
      \nu_\alpha\right] }} \ ,
\end{equation}

\noindent
where $\nu_\alpha$ are the depolarization factors related to the
principal values of $\hat{Q}$ by

\begin{equation}
\nu_\alpha = 1/3 - Q_\alpha
\end{equation}

\noindent
In the case of spherical symmetry ($e=0$), $Q_\alpha = 0$ and
$\nu_\alpha=1/3$. For an ellipsoid of nonzero eccentricity, the
depolarization factors become different from $1/3$. Thus, for example,
if $e=1$, we have $\nu_1=\nu_2=0, \ \nu_3=1$ for an oblate ellipsoid
(infinitely thin circular disk) or $\nu_1=\nu_2=1/2, \ \nu_3=0$ for a
prolate ellipsoid (infinitely thin needle).  The anisotropy factor $S$
can be defined as dispersion of the depolarization factors:

\begin{equation}
\label{S_def}
S^2 = \langle \nu_\alpha^2 \rangle - \langle \nu_\alpha \rangle^2 \ .
\end{equation}

\noindent
Obviously, this parameter is zero for a sphere and positive for any
ellipsoid of nonzero eccentricity. In particular, for the infinitely
thin needle, $S=1/3\sqrt{2}$ and for an infinitely thin circular disk,
$S=\sqrt{2}/3$. The latter is the maximum possible value for $S$ given the
constraint $\sum_\alpha \nu_\alpha=1$.

Now we extend the definition of the depolarization tensor to include
particles of arbitrary shape. Namely, for an arbitrary system
occupying some volume $V$, we define

\begin{equation}
\label{nu_def}
\hat{\nu} ({\bf r}) =
{1 \over 3}\hat{I} - \int_V \hat{G}_0({\bf r}, {\bf
r}^{\prime})d^3r^{\prime} \ ,
\end{equation}

\noindent
where $\hat{I}$ is the unity tensor. If $V$ is of general shape, the
result of integration in the right-hand side of (\ref{nu_def}) is
position-dependent.  Therefore, the tensor $\hat{\nu}({\bf r})$
depends on the point ${\bf r}$ it is evaluated at and is referred to
here as {\it local}.  Similarly to the case of ellipsoids, this tensor
can be diagonalized. Then we can use the principal values
$\nu_\alpha({\bf r})$ to calculate the anisotropy factor according to
(\ref{S_def}).

In this paper we consider aggregates of (possibly, polydisperse)
spheres whose centers are located at points ${\bf r}_i$ and radii are
denoted by $a_i$.  In this case, the expression for $\hat{\nu}({\bf
  r})$ is simplified. We use

\begin{equation}
\int_{\vert{\bf r}^{\prime} - {\bf r}_i\vert<a_i} \hat{\bf G}_0({\bf
  r},{\bf r}^{\prime}) d^3 r^{\prime} =
\left\{
\begin{array}{ll}
v_i \hat{\bf G}_0({\bf r},{\bf r}_i) \ , &  {\rm if} \
\vert{\bf r} - {\bf r}_i\vert > a_i \\
0 \ , & {\rm if} \
\vert{\bf r} - {\bf r}_i\vert < a_i
\end{array}
\right.
\end{equation}

\noindent
to obtain

\begin{equation}
\hat{\nu}_i \equiv \hat{\nu}({\bf r}_i) =
{1 \over 3}\hat{I} - \sum_{j\neq i} v_j
\hat{G}_0({\bf r}_i, {\bf r}_j) \ .
\end{equation}

\noindent
Here $v_i=4\pi a_i^3/3$ is the volume of the $i$-th sphere and
the components of $\hat{G}_0({\bf r}_i, {\bf r}_j)$ are given by

\begin{equation}
\label{G_0_def}
\left(G_0({\bf r}_i, {\bf r}_j)\right)_{\alpha\beta} =
\frac{\delta_{\alpha\beta}-3n^{(ij)}_{\alpha}n^{(ij)}_{\beta}}{r^3_{ij}}
\ ,
\end{equation}

\noindent
where ${\bf r}_{ij}={\bf r}_i - {\bf r}_j$ and ${\bf n}^{(ij)}={\bf
  r}_{ij}/r_{ij}$.

Diagonalization of the tensor $\hat{\nu}_i$ and calculation of the
dispersion of its principal values gives the local anisotropy factor
$S_i$. This parameter quantifies the degree of anisotropy of the local
environment of the $i$-th sphere.

A few notes about the introduced definition must be made. First, the
principal value $\nu_\alpha$ obtained as described above are purely
geometrical characteristics of an object. They are related to the
Bergman-Milton spectral parameters~\cite{bergman_pr} only in the
special case of ellipsoidal (more generally, spheroidal) shape of $V$.
Obtaining the Bergman-Milton spectral parameters requires
diagonalization of the integral operator $W$ with the kernel $G_0({\bf
  r},{\bf r}^{\prime})$, ${\bf r}, {\bf r}^{\prime} \in V$. This is a
much more complicated task than diagonalization of the tensor
$\hat{Q}({\bf r})=\int_V G_0({\bf r},{\bf r}^{\prime})d^3r^{\prime}$
at a given point ${\bf r}$. In particular, $\hat{Q}({\bf r})$ is
three-dimensional, while $W$ is infinite-dimensional. Correspondingly,
the number of Bergman-Milton parameters is infinite (although only
three of them have non-zero oscillator strengths in the case of
spheroids), while the tensor $\hat{\nu}({\bf r})$ has only three principal
values.  Second, the principal values $\nu_{\alpha}({\bf r})$ are not
constrained, in general, by the conditions $0 \leq \nu_\alpha \leq 1$
and $\sum_\alpha \nu_\alpha = 1$. This also distinguishes them from
the Bergman-Milton spectral parameters. Next, the parameter $S_i$
depends on the coordinates of all nanoparticles in the aggregate with
$j\neq i$. However, due to the fast cubic decay of the near-field
component of the dipole radiation field, the neighbors within few
coordinate spheres of the $i$-th site give the largest input to $S_i$.
This justifies the locality of $S_i$, as it only weakly depends on the
large scale structure. This statement needs to be qualified in
aggregates large enough so that interaction in the far-zone becomes
important. Even without account of retardation, the locality of $S_i$
can be violated in aggregates with the fractal dimension close to $3$
(or in random non-fractal composites), due the logarithmic divergence
of the integral $\int r^{-3} d^3r$ at infinity. We do not expect these
effects to be important in most aggregates of practical interest with
the fractal dimension in the range $D<2.7$ and do not consider them in
this paper.

Finally, the introduced parameter is not sensitive to the wavelength
and electromagnetic properties of the scattering material. Therefore,
we do not expect it to be a good indicator of local electromagnetic
response at all wavelengths. It is also independent of the incident
polarization. A possible definition of a polarization-sensitive
anisotropy factor is

\begin{equation}
\label{S_E0_1}
S_i({\bf E}_0) = \frac{{\bf E}_0^* \cdot \hat{\nu}_i {\bf E_0}}{\vert
  {\bf E}_0 \vert^2} \ . 
\end{equation}

\noindent
Another possible definition is

\begin{equation}
\label{S_E0_2}
S_i^2({\bf E}_0) = \frac{\vert \hat{\nu}_i {\bf E_0}\vert^2}{\vert
  {\bf E}_0 \vert^2} \ . 
\end{equation}

\noindent
Note that the definitions (\ref{S_E0_1}),(\ref{S_E0_2}) are not used
in this paper.

\section{Results: simulations of geometrical properties}
\label{sec:sim}

Since the unique electromagnetic properties of colloid aggregates are
often attributed to their fractal structure, we have studied
computer-generated aggregates with various fractal dimensions.  We
have generated quasi-random off-lattice aggregates with varying
fractal dimension $D$ using the algorithm described
in~\cite{markel_04_3}. This algorithm simulates the stochastic
dynamics of individual nanoparticles and sub-aggregates in a solution
with the account of random (Brownian) forces, as well as deterministic
interparticle (the Van-der-Waals and Coulomb) and external potentials.
Discrete Newtonian mechanics was implemented with a sufficiently small
time step, such that the spatial translation of any particle (sphere)
at each step is much smaller than its diameter. Rotation of aggregates
was taken into account. We have used both monodisperse ($a_i={\rm
  const}$) and polydisperse nanospheres ($a_i$ were randomly
distributed according to the Poisson distribution). The fractal
dimension of obtained aggregates was tuned in the interval $1.7<D<3.0$
by varying the initial density of spheres prior to the aggregation
process. The numerical value of $D$ was calculated from the linear
regression of the pair density-density correlation function which, in
the intermediate asymptote region, has the scaling form $g(r) \propto
r^{D-3}$.

The aggregation was simulated in a cubic volume with elastically
reflecting boundaries. In the limit of low initial concentration of
particles and the size of the cubic cell of $\sim 200a$ or more, the
obtained aggregates have the typical fractal dimension $D \approx
1.7$. When the initial concentration increases, $D$ approaches the
trivial limit $D=3$. As a graphical illustration of generated
fractals, we show in Fig.~1 a large aggregate and values of the local
anisotropy factor $S$ at some selected sites.\\

%\begin{figure}[h!]
\psfig{file=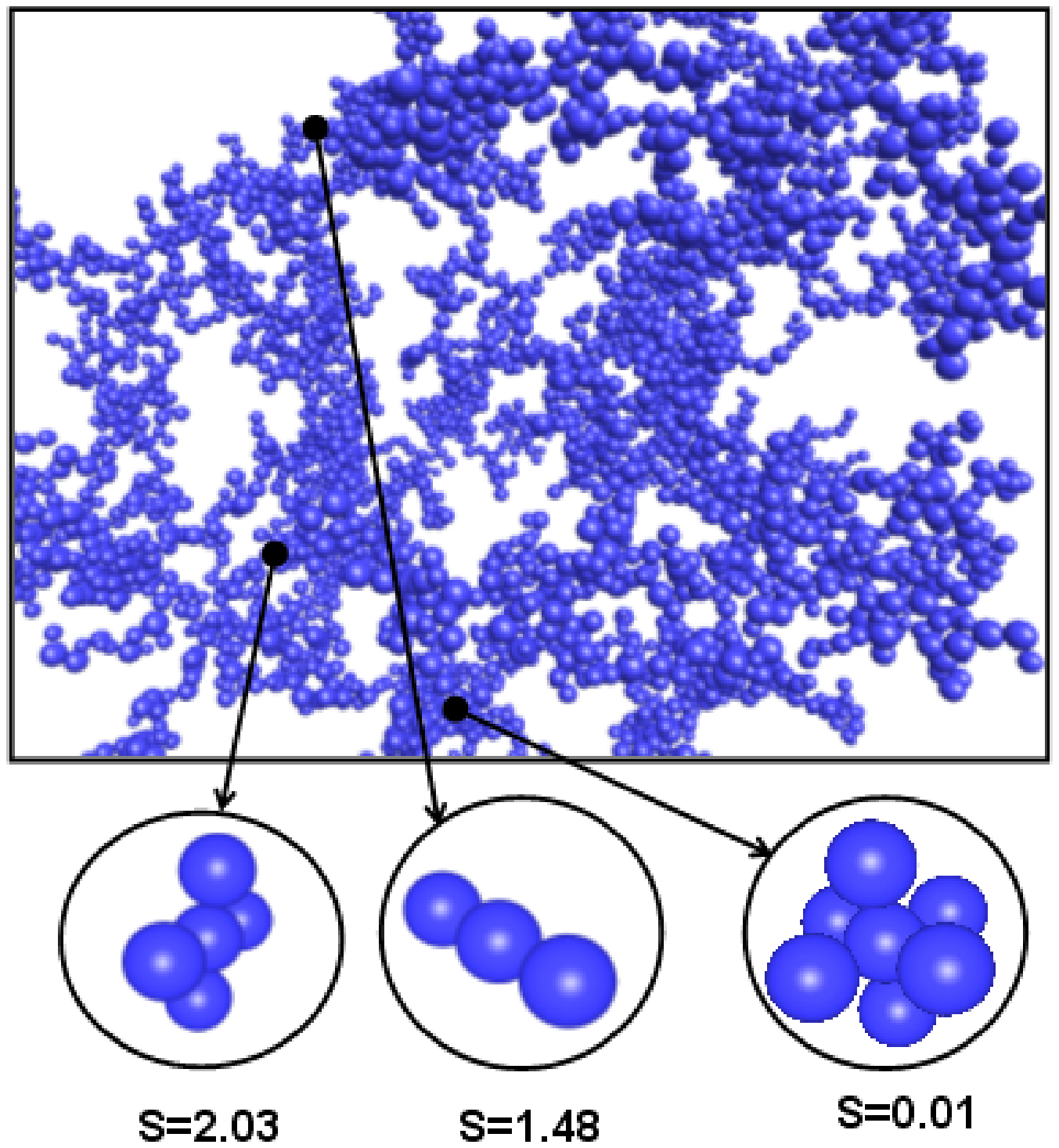,width=8.2cm}
{\small \noindent Fig.~1. Local anisotropy factor $S$ for selected sites
  in a large aggregate with $N=5000$ and $D \approx 1.8$.}\\
%\end{figure}

We start with a discussion of results for monodisperse aggregates,
i.e., for aggregates built of identical spheres. In Fig.~2, we
illustrate the dependence of the average (over individual particles in
an aggregate) value of the $S$ on fractal dimension $D$.  Aggregates
with $1.70<D<2.25$ are characterized by moderate average values of
local anisotropy factor, almost independently of $D$. We can argue
that such aggregates differ only on large scale but have similar local
structure. In other words, the local environment of each particle is,
on average, the same, independently of $D$. As $D$ approaches the
critical value $D=3$, the local anisotropy factor quickly drops. This
corresponds to the fact that trivial (non-fractal) aggregates are
characterized by almost isotropic local environment and relatively
weak fluctuations of density.  We have also calculated the average $S$
for two types of lattice aggregates traditionally used in
electrodynamic calculations. The results are shown by centered symbols
in Fig.~2.\\

%\begin{figure}[h!]
  \psfig{file=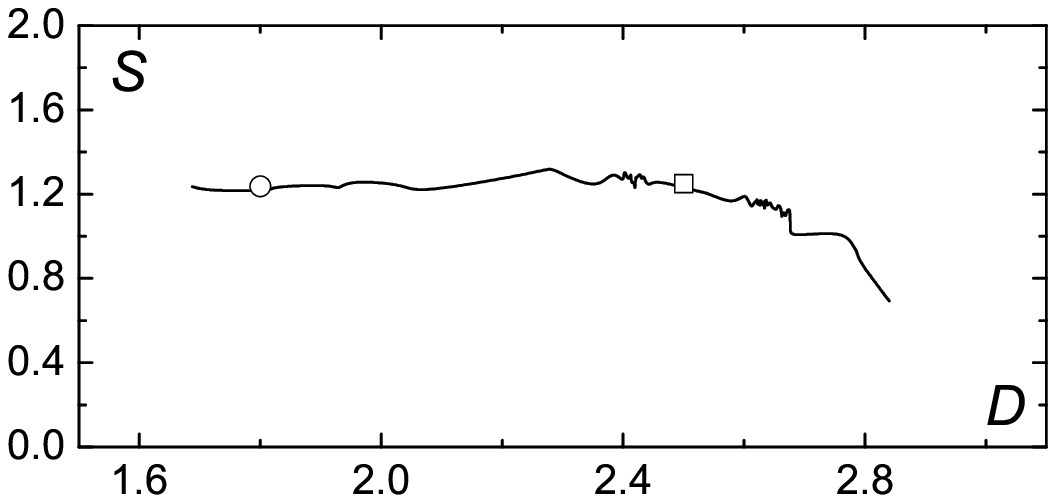,width=8.2cm}
{\small \noindent Fig.~2. Average local anisotropy factor $S$ vs fractal
  dimension $D$ for $121$ random aggregates with $N=800$ in each.
  Numerical value of $D$ was computed separately for each aggregate
  from linear regression of the density-density correlation function
  in the intermediate asymptote region. The open circle shows the
  average values of $S$ for lattice Meakin
  aggregates~\cite{meakin_83_1} ($D \approx 1.8$) and the open square
  shows the same value for a set of Witten-Sander
  aggregates~\cite{witten_81_1} ($D \approx 2.5$).}\\
%\end{figure}

Real colloid aggregates are strongly polydisperse. Typically, they
contain particles of sizes ranging from $5{\rm nm}$ to $30{\rm
  nm}$~\cite{kreibig_book_95,heard_83_1}. We have investigated the
dependence $S(D)$ for several ensembles of polydisperse aggregates
with different ratios of the maximum and minimum sphere radiuses,
$a_{\rm max}$ and $a_{\rm min}$. We have used a discrete Poisson
distribution of particle sizes with the number of samples equal to
$11$ [the discrete step in particle size was $\Delta a = (a_{\rm max}
- a_{\rm min})/10$]. The dependence of local anisotropy factor on the
fractal dimensions $D$ is shown in Fig.~3. Note that no signifact
effect doe to the polidispersity was found.

It is interesting to note that the average local anisotropy factor
does not depend on the distance of a given site from the center of
mass of the aggregate. This is illustrated in Fig.~4. Here we plot the
value of $S$ averaged over all nanospheres within a spherical shell
drawn arond the center of mass of the aggregate as a function of the
shell radius (see figure captions for more detail).

In Fig.~5, we also plot the fraction of particles in an
aggregate with local anisotropy factor exceeding 60\% of the maximum
value for that aggregate as a function of fractal dimension. It can be
seen that in typical aggregates with fractal dimensions of practical
interest, only small fraction of particles is placed in highly
anisotropic environment. In Fig.~6, an example of an
aggregate is 

%\begin{figure}[h!]
\psfig{file=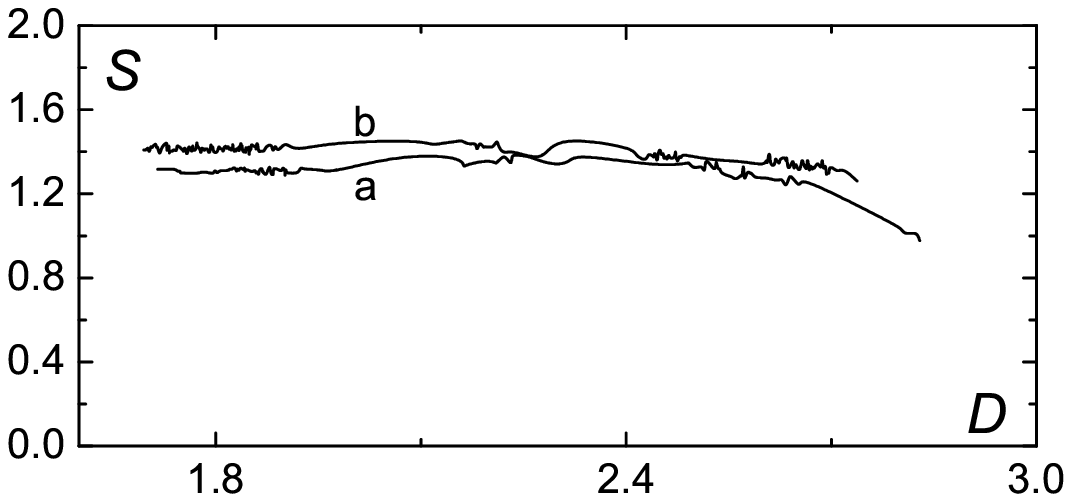,width=8.2cm}
{\noindent \small Fig.~3. Local anisotropy factor $S$
  factor vs fractal dimensions $D$ for polydisperse aggregates with
  $N=800$ and Poisson particle size distribution. The ratio of the
  maximum and minimum particle radii is $a_{\rm max}/a_{\rm min}=2$
  (153 random aggregates) for curve (a) and $a_{\rm max}/a_{\rm
    min}=3$ (297 random aggregates) for curve (b). Numerical value of
  $D$ was computed separately for each aggregate from linear
  regression of the density-density correlation function in the
  intermediate asymptote region.}
%\end{figure}

%\begin{figure}[h!]
\psfig{file=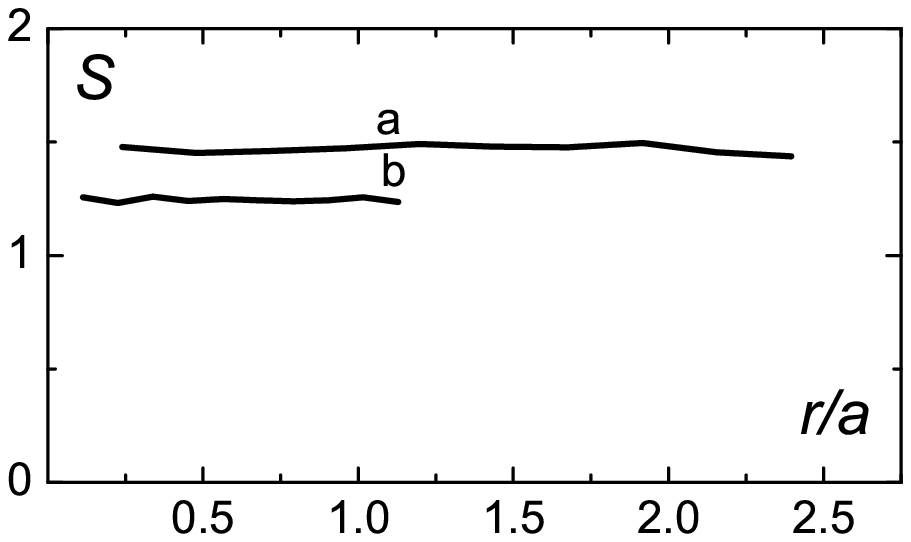,width=8.2cm}
{\noindent \small Fig.~4. Local anisotropy
  factor $S$ vs the relative distance to the center of mass of an
  aggregate, $r/a$. The hystogram is built with the step $R_g/10$,
  where $R_g$ is the gyration radius of the aggregate, and $S$ was
  averaged over all particles located within $10$ spherical shells
  drawn around the aggregate's center of mass for $N=10,000$ (a) and
  $N=3,000$ (b).}
%\end{figure}

%\begin{figure}[h!]
\psfig{file=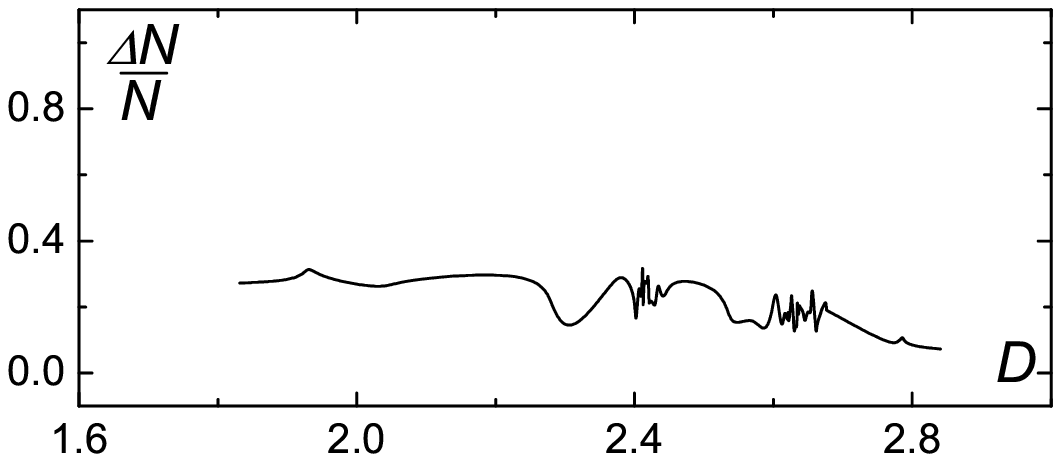,width=8.2cm}
{\noindent \small Fig.~5. Dependence of the
  fraction $\Delta N / N$ of sites in an monodisperse aggregate with
  the value of local anisotropy factor exceeding 60\% of its maximum
  value for the same aggregate; $N=800$.}
%\end{figure}

%\begin{figure}[h!]
\psfig{file=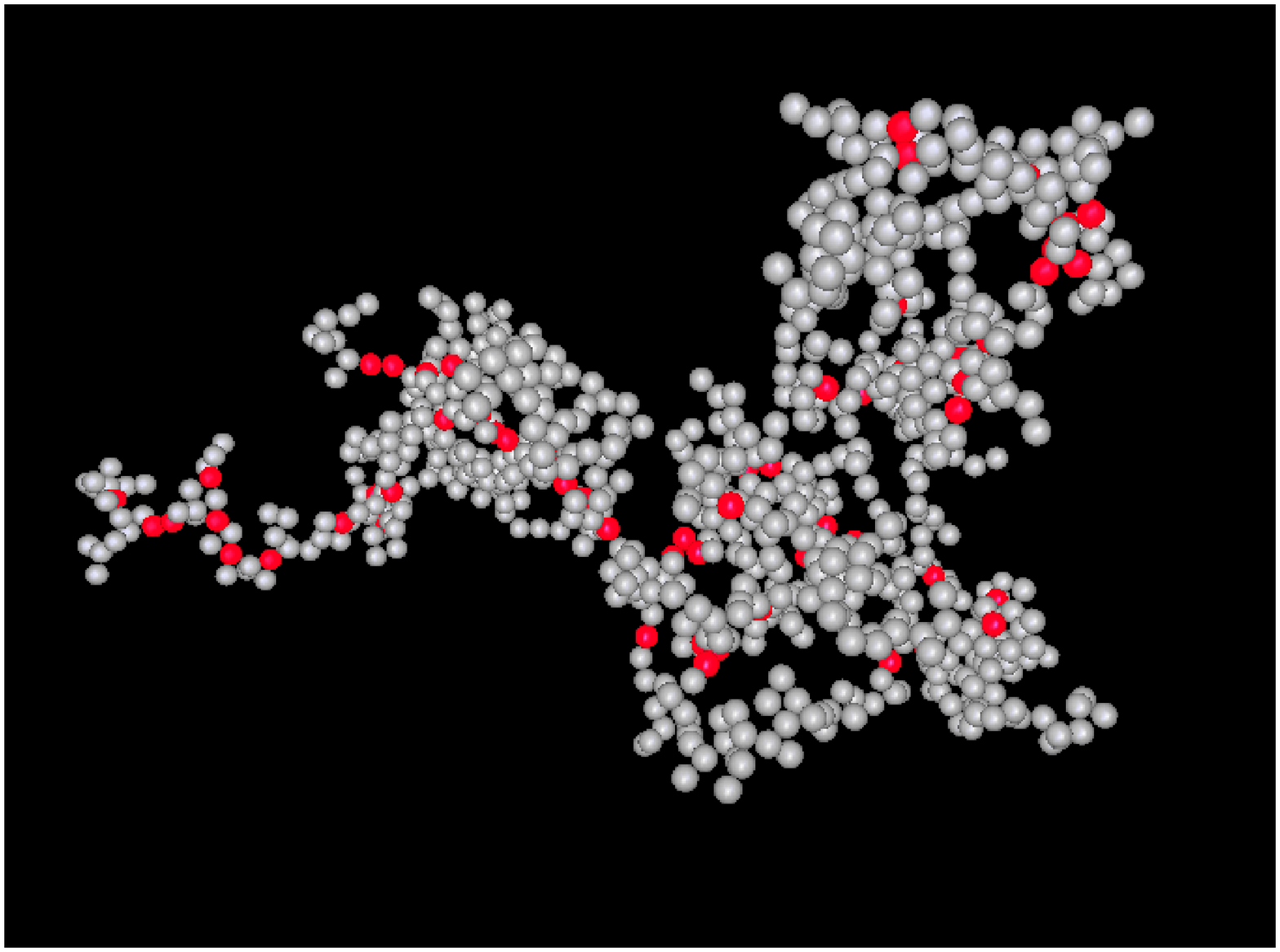,width=8.2cm}
{\noindent \small Fig.~6. Sites in a fractal aggregate ($D\approx 1.8$,
  $N=800$) with relative values of the local anisotropy factor
  exceeding 80\% of the maximum value for the same aggregate, $S_{\rm
    max}=2.29$.}\\ \\
%\end{figure}

%\begin{figure}[h!]
\centerline{\psfig{file=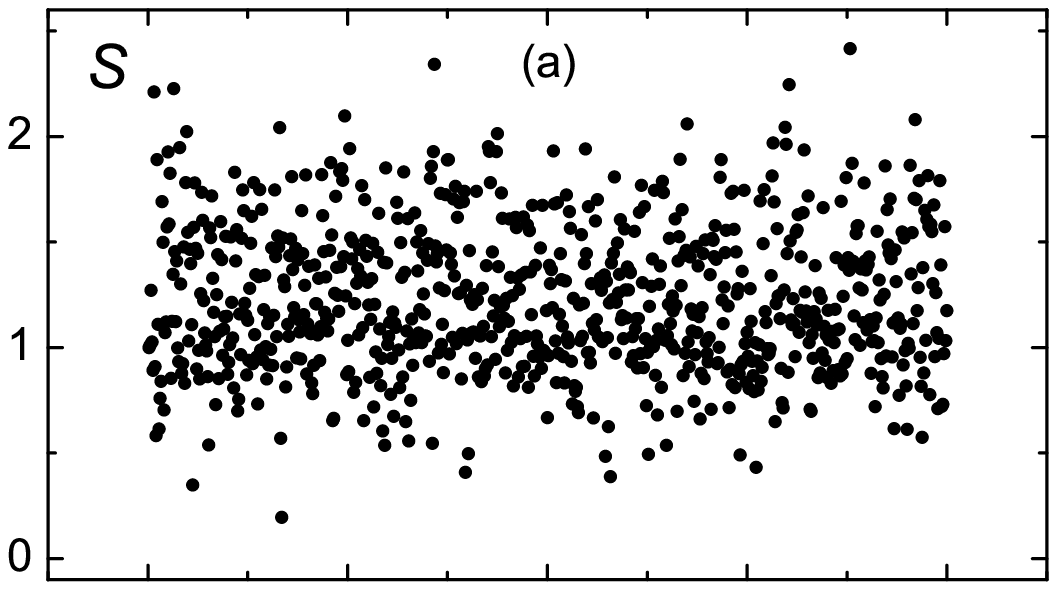,width=8.2cm,bbllx=0bp,bblly=30bp,bburx=355bp,bbury=200bp,clip=}}
\centerline{\psfig{file=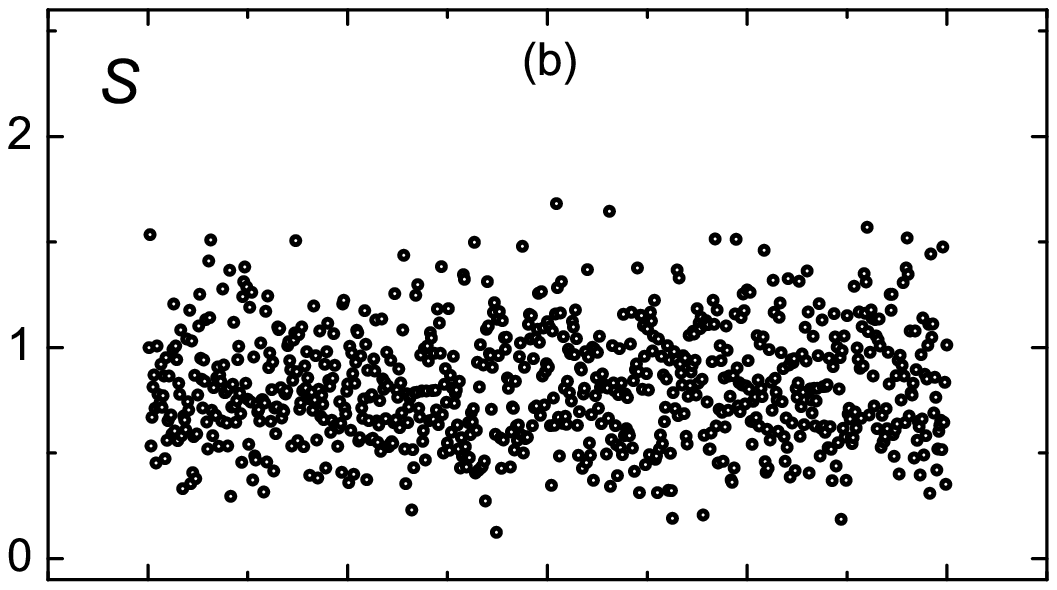,width=8.2cm,bbllx=0bp,bblly=30bp,bburx=355bp,bbury=200bp,clip=}}
\centerline{\psfig{file=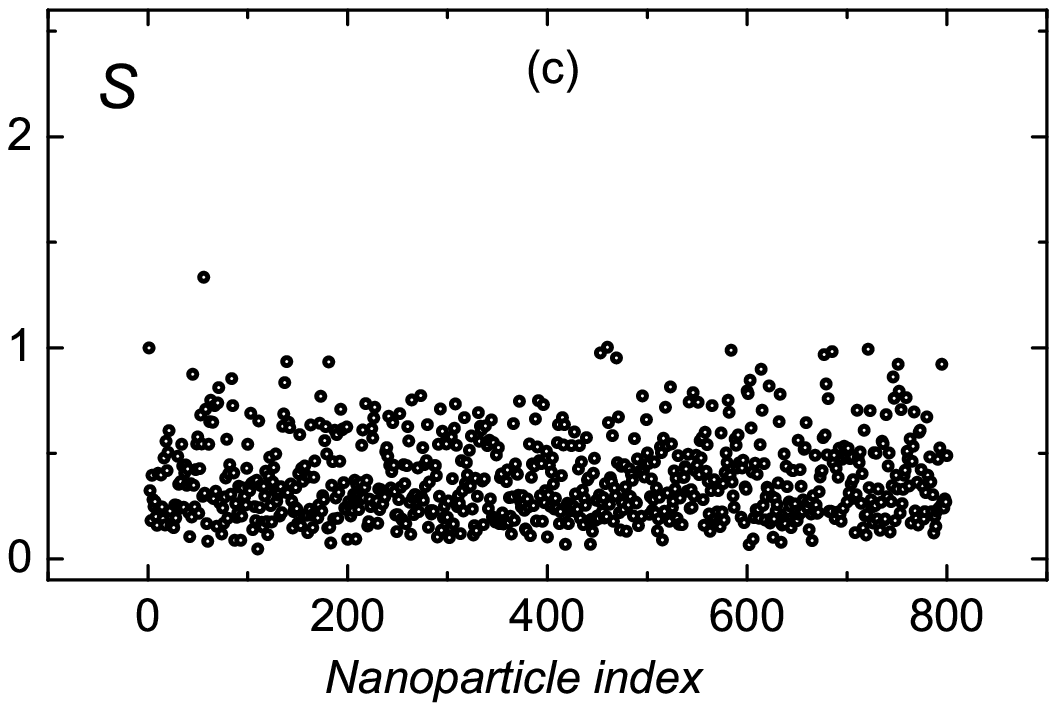,width=8.2cm,bbllx=0bp,bblly=0bp,bburx=355bp,bbury=220bp,clip=}}
{\noindent \small Fig.~7. Local anisotropy factor $S$ for different
  particles in a fractal aggregate with $N=800$ and $D\approx 1.8$ (a)
  compared to those in random gas of identical hard spheres.Average
  distance between centers of two nearest neighbor spheres $R_{\rm
    nn}/a=2.12$ (b) and $R_{\rm nn}/a=2.90$ (c).}
%\end{figure}

\newpage
\noindent
shown with the sites of relatively high local anisotropy
emphasized by different color (shade of grey).

Finally, we compare the local anisotropy factors for all particles of
monodisperse fractal aggregate and non-fractal random gas of hard
spheres ($N=800$ in both cases). In Fig.~7a we plot these
quantities for a fractal aggregate with $D\approx 1.8$.  All local
anisotropy factors $S_i$ are shown for $i=1,\ldots,800$. In
Fig.~7b,c, the same quantities are plotted for a random gas
of identical hard spheres of radius $a$ distributed in a volume with
the density corresponding to the average distance between the centers
of nearest neighbor spheres equal to $R_{\rm nn}$; value of the ratio
$R_{\rm nn}/a$ are indicated in the figure caption. It can be seen
that the fractal aggregate contains sites with much higher values of
local anisotropy factor than random gas. As one could expect, the
local anisotropy factors become smaller when the density of random gas
decreases.  However, a fractal aggregate, although has zero asymptotic
density in the limit $N\rightarrow \infty$, always retains
approximately constant fraction of sites with relatively high local
anisotropy.\\

%\begin{figure}[h!]
\centerline{\psfig{file=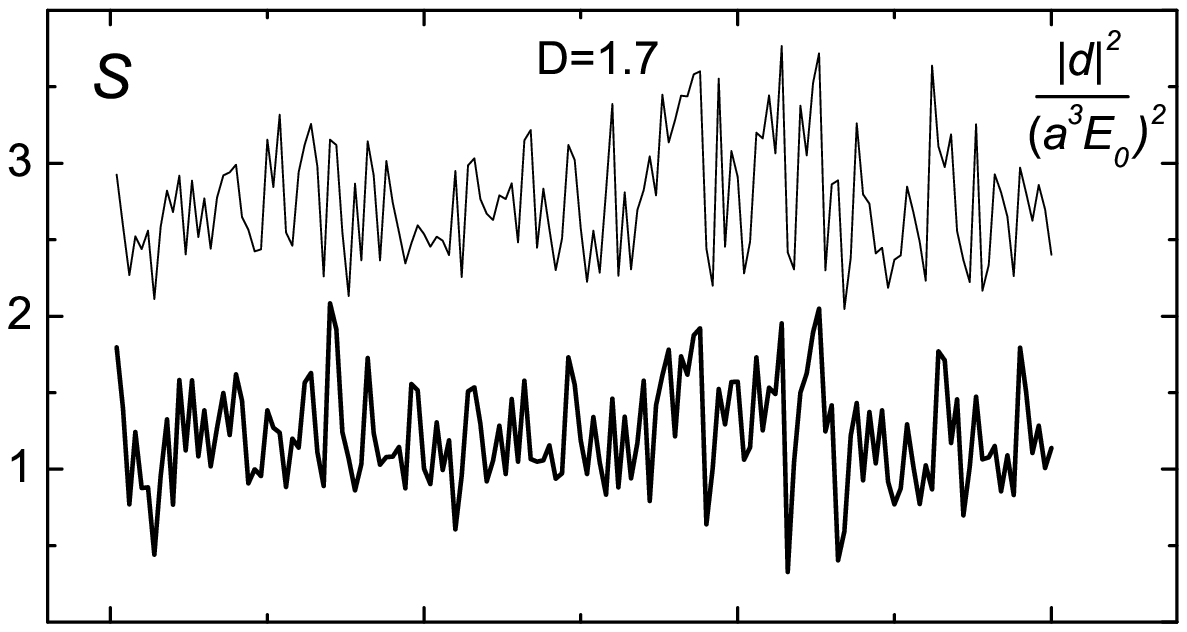,width=8.2cm,height=4.00cm,bbllx=0bp,bblly=30bp,bburx=375bp,bbury=220bp,clip=}}
\centerline{\psfig{file=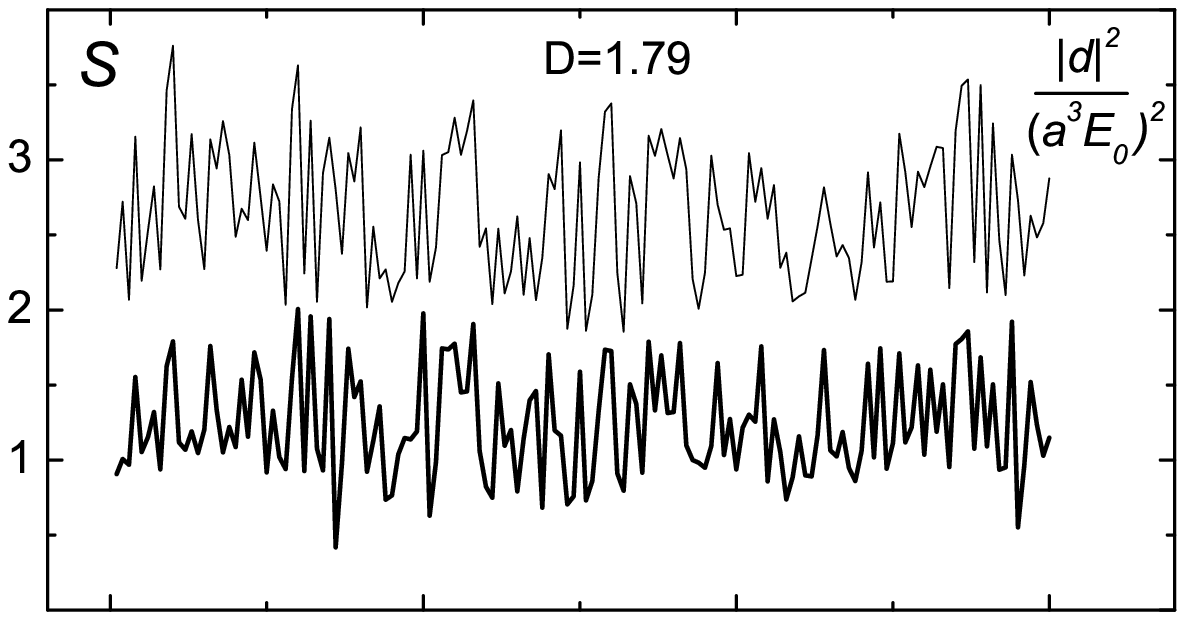,width=8.2cm,height=3.75cm,bbllx=0bp,bblly=30bp,bburx=375bp,bbury=207bp,clip=}}
\centerline{\psfig{file=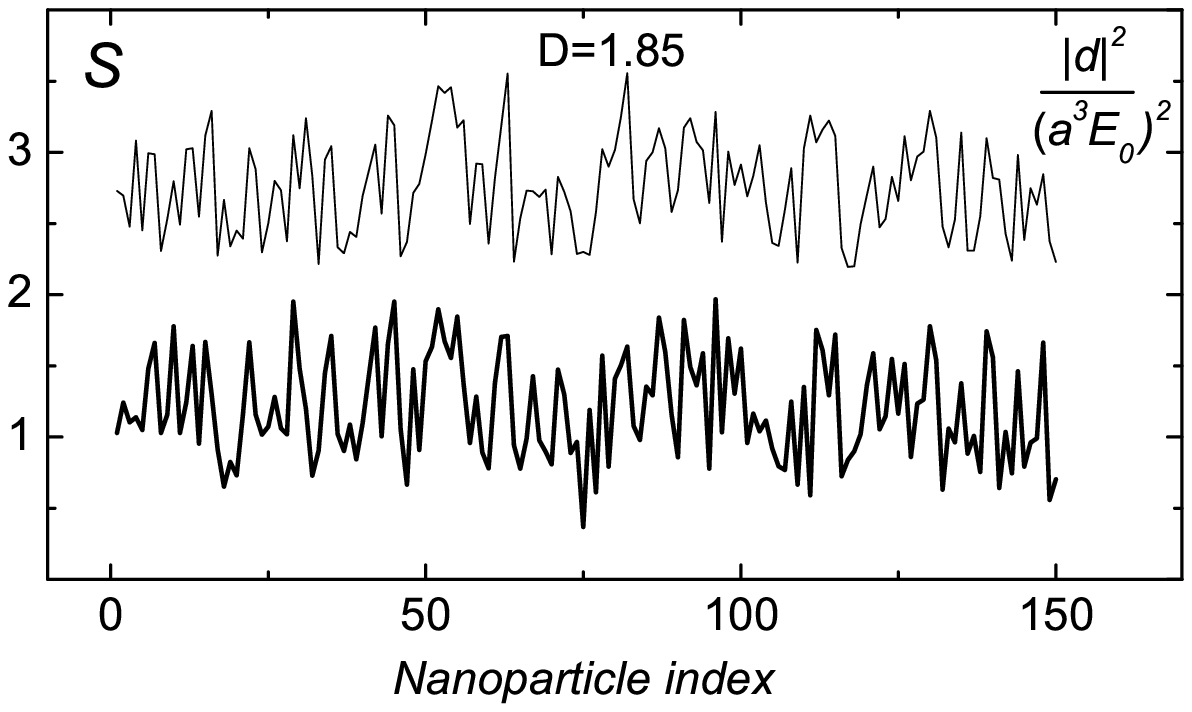,width=8.2cm,height=4.76cm,bbllx=0bp,bblly=10bp,bburx=375bp,bbury=220bp,clip=}}
\vspace*{-0.25cm}
{\noindent \small Fig.~8. Local anisotropy factor ($S_i$) (thick
  line) and local dipole moments squared $\vert {\bf d}_i \vert^2$
  (thin line) for different particles in a monodisperse aggregate with
  $N=150$ and fractal dimension $D=1.70$, $D=1.79$, and $D=1.85$,
  computed at the wavelength $\lambda = 703{\rm nm}$.}
%\end{figure}

%\begin{figure}[h!]
\psfig{file=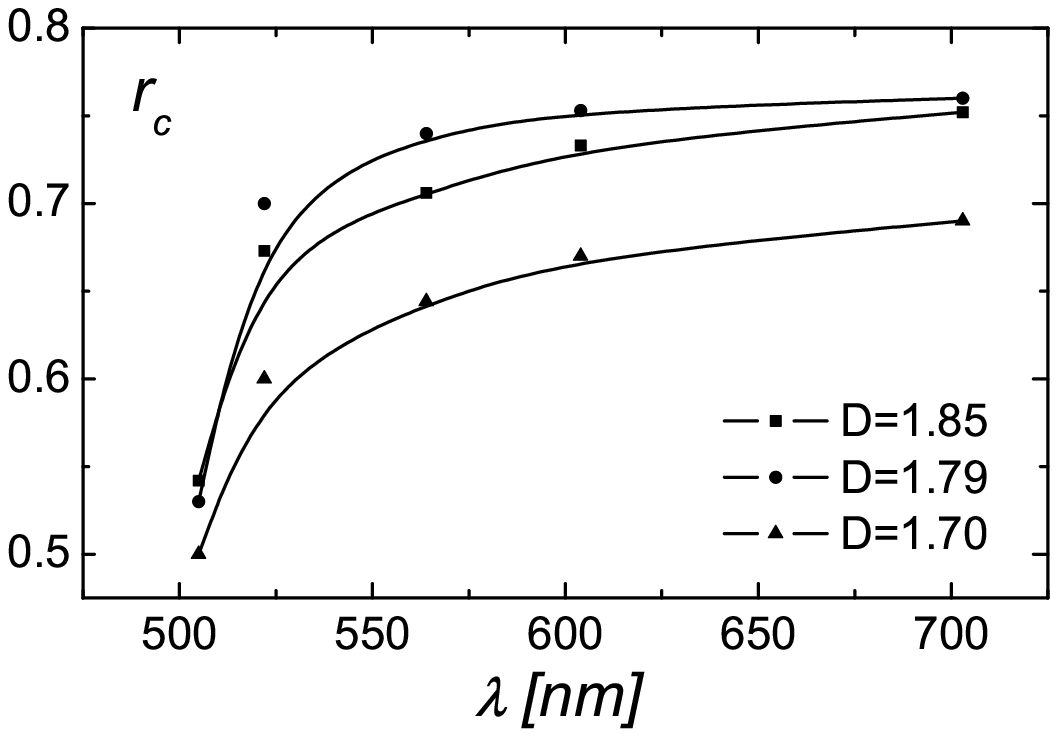,width=8.2cm}
{\noindent \small Fig.~9. Correlation between $S_i$ and $\vert {\bf d}_i
  \vert^2$ as a function of wavelength for monodisperse aggregates
  with $N=150$ and fractal dimension $D=1.70$, $D=1.79$, and
  $D=1.85$.}
%\end{figure}
\section{Comparison of structural and electrodynamic
  properties of fractal nanoaggregates}
\label{sec:opt}

The main idea of this paper is that there is certain correlation
between local structure and local electromagnetic fields in fractal
nanoaggregates. This assumption is confirmed by numerical simulations
presented in this Section.

We have computed optical responses of aggregates of nanospheres using
the method of coupled multipoles~\cite{mackowski_95_1,markel_04_3}.
Calculations were performed for monodisperse aggregates built of
$N=150$ silver nanospheres of constant radius $a=5{\rm nm}$ and placed
in vacuum. To facilitate convergence with the maximum order of
multipoles included, we have introduced a surface layer of thickness
$h=0.05a$. The dielectric constant of the layer was chosen to be the
same as that of the vacuum, $\epsilon=1$. We have used experimental
values of the optical constants of silver~\cite{johnson_72_1} with
finite-size corrections according to~\cite{markel_96_1}. The maximum
order of the VSHs utilized in the results shown below was $L=8$. The
convergence was verified by control calculations with $L=16$. We note
that much larger values of $L$ are required for nanospheres in exact
contact ($h=0$) and that the number of the coupled-multipole equations
(with complex coefficients) which must be solved to compute the
optical responses is equal to $NL(L+2)$.

In Fig.~8 we plot the quantities $S_i$ and $\langle \vert
{\bf d}_i \vert^2 \rangle /(a^3E_0)^2$ for three aggregates with
fractal dimensions $D\approx 1.70$, $D\approx 1.79$ and $D\approx
1.85$, computed at $\lambda=703{\rm nm}$. Here $\langle \vert {\bf
  d}_i \vert^2 \rangle$ is the square of the dipole moment of $i$-th
nanosphere averaged over three orthogonal polarizations of the
external field.  Visual correlation of the two curves is quite
apparent. For a more quantitative estimate, we have computed the
correlation coefficient $r_c(S, \langle \vert {\bf d}\vert^2
\rangle)$. The dependence of $r_c$ on the wavelength is shown in
Fig.~9 for three values of fractal dimension and different
wavelengths. The maximum degree of correlation is achieved for
$\lambda=703{\rm nm}$ ($0.69<r_c<0.76$). The value of $r_c$ decreases
monotonously for smaller wavelengths and is in the interval
$0.49<r_c<0.54$ when $\lambda = 505{\rm nm}$. Note that the
correlation coefficient is expected to increase towards unity in the
spectral region $\lambda>700{\rm nm}$. Also, even stronger correlation
is expected is a polarization-dependent definition of the local
anisotropy factor is used, such as (\ref{S_E0_1}) or (\ref{S_E0_2}).
Validating these hypothesis will be the subject of future work.

\section{Summary}
\label{sec:sum}

In this paper we have investigated the statistical correlation between
the local geometrical structure and local electromagnetic responses in
fractal aggregates of nanoparticles. We have used a realistic
aggregation model which allows computer generation of quasi random
aggregates of variable fractal dimension in the interval $1.7<D<3.0$.
Electromagnetic calculations were carried out using the method of
coupled multipoles, i.e., beyond the dipole approximation.

We have found that the local anisotropy factor $S$ introduced in
Section~\ref{sec:lap_def} is strongly correlated with the local
electromagnetic response. For aggregates built of high-quality
plasmonic materials, the degree of such correlation tends to increase
with the wavelength. The correlation coefficient between the squared
dipole moment of a given nanoparticle in an aggregate and a purely
geometrical parameter (local anisotropy factor) reaches the value of
$\approx 0.75$ for $\lambda=700{\rm nm}$. We expect that this
correlation can become even larger if a properly-defined
polarization-dependent local anisotropy factor is used and at larger
wavelengths.

The introduced parameter $S$ is a universal geometrical characteristic
which can be used for analyzing various complicated aggregates and
composites without explicit solution of the electromagnetic problem.
The discovered strong correlation suggests that, at least in
aggregates which are small compared to the wavelength, the large-scale
geometry does not play a significant role. Note that in the IR
spectral region, subwavelength aggregates can still be built of
hundreds or even thousands of nanospheres. The IR spectral region is
of special interest because of the very low Ohmic losses in silver and
other noble metals. Correspondingly, heterogeneous nanostructures are
known to exhibit optical resonances of very high quality. This, in
turn, results in giant amplification of local optical responses. The
latter phenomenon is currently being actively researched in ordered
nanostructures, including self-similar chains of nanospheres
(nanolenses)~\cite{li-kuiru_03_1} and long chains of similar
nanoparticles~\cite{zou_05_1}. Rigorous numerical simulations in
random nanoaggregates are still difficult due to the high
computational complexity of the associated electromagnetic problem.
The introduced parameter $S$ and the discovered correlation of this
parameter with local electromagnetic field allows one to make
qualitative predictions about the sites where the electromagnetic
energy is localized by very simple means, e.g., without solving the
electromagnetic problem.

\section*{Acknowledgements}

This research was supported by the Russian Foundation for Basic
Research, Grant 05-03-32642, and by the Presidium of Russian Academy
of Sciences under the Grant 8.1 "Basic Problems of Physics and
Chemistry of Nanosize Systems and Nanomaterial". 

\bibliographystyle{prsty}
\bibliography{abbrevplain,article,book,local}

\end{document}